\documentclass[letterpaper, 10 pt, conference]{Attachment/ieeeconf}

\IEEEoverridecommandlockouts                              
                                                          
\usepackage[utf8]{inputenc}
\usepackage{amsmath}
\usepackage{mathtools}
\usepackage{amsfonts}
\usepackage{xcolor}
\usepackage[english]{babel}
\usepackage{stackengine}
\usepackage{amssymb}
\usepackage{mathrsfs}
\usepackage{algorithm, float}
\usepackage{algorithmic}
\usepackage{bbm}
\usepackage{dsfont}
\usepackage{booktabs} 
\usepackage{caption}
\usepackage{subcaption}

\newtheorem{The}{Theorem}
\newtheorem{ass}{Assumption}
\newtheorem{lemma}{Lemma}
\newtheorem{remark}{Remark}
\newtheorem{definition}{Definition}
\newtheorem{Pro}{Problem}

\newtheorem{corollary}{Corollary}

\def\Var{{\textrm{Var}}\,}

\DeclareMathOperator{\Tr}{Tr}

\DeclareMathOperator*{\argmin}{arg\,min}
\newcommand{\vect}[1]{\mathbf{#1}}
\DeclareMathOperator{\E}{\mathbb{E}}

\title{Path Integral Methods with Stochastic Control Barrier Functions}

\author{Chuyuan Tao$^{\dagger}$, Hyung-Jin Yoon$^{*}$, Hunmin Kim$^{\dagger}$, Naira Hovakimyan$^{\dagger}$, and Petros Voulgaris$^{*}$
\thanks{This research is supported by NSF CPS \#1932529, NSF CMMI \#1663460, UNR internal funding, and UIUC STII-21-06.}
\thanks{$^{\dagger}$Chuyuan Tao, Hunmin Kim, and Naira Hovakimyan are with the Department of Mechanical Science and Engineering, University of Illinois at Urbana-Champaign, USA.
{\tt\small  \{chuyuan2, hunmin, nhovakim\}@illinois.edu}}%
\thanks{$^{*}$Hyung-Jin Yoon and Petros Voulgaris are with the Department of Mechanical Engineering, University of Nevada, Reno, USA.
{\tt\small  \{hyungjiny, pvoulgaris\}@unr.edu }}
}

\begin{document}

\maketitle
\begin{abstract}
Safe control designs for robotic systems remain challenging because of the difficulties of explicitly solving optimal control with nonlinear dynamics perturbed by stochastic noise. However, recent technological advances in computing devices enable online optimization or sampling-based methods to solve control problems. For example, Control Barrier Functions (CBFs), a Lyapunov-like control algorithm, have been proposed to numerically solve convex optimizations that determine control input to stay in the safe set. Model Predictive Path Integral (MPPI) uses forward sampling of stochastic differential equations to solve optimal control problems online. Both control algorithms are widely used for nonlinear systems because they avoid calculating the derivatives of the nonlinear dynamic function. In this paper, we utilize Stochastic Control Barrier Functions (SCBFs) constraints to limit sample regions in the sample-based algorithm, ensuring safety in a probabilistic sense and improving sample efficiency with a stochastic differential equation. We provide a sampling complexity analysis for the required sample size of our algorithm and show that our algorithm needs fewer samples than the original MPPI algorithm does. Finally, we apply our algorithm to a path planning problem in a cluttered environment and compare the performance of the algorithms.
\end{abstract}

\section{Introduction}
Safety verification is crucial when applying control algorithms to robotic systems in the presence of uncertainties. Failure to ensure safety may cause severe damage to robots, properties, and people nearby. There are numerous existing papers that aim to guarantee safety. To list a few, the reachable sets method in~\cite{girard2006efficient} designs control trajectories while monitoring whether the reachable sets violate safety constraints. The barrier certificate method in~\cite{prajna2004safety} uses the dynamics and the certificate functions associated with the safety constraint inequality to ensure safety. The robust model predictive method (MPC) in~\cite{bemporad1999robust, lofberg2003approximations} employs  min-max optimization  to improve the robustness against disturbances. Another seminal safety verification method is the Control barrier function (CBF), which utilizes a Lyapunov-like function (i.e., the reciprocal CBF) to guarantee that the control output is forward invariant in a defined safe set~\cite{ames2019control}.
Since the CBF can be augmented to a nominal control by solving a quadratic optimization program which can be implemented online, the CBF method is widely used in real-time robotic systems with extensions~\cite{ames2019control}. For example, the authors in~\cite{majd2021safe} combine the CBF with the Rapidly exploring Random Tree (RRT) algorithm to robustly satisfy the collision constraints in real-time. Also, the authors in~\cite{cosner2021measurement} use \emph{Backup Sets} to find admissible inputs and construct Measurement-Robust CBF that provides a margin to the robustness against measurement noise.


There are two popular ways to augment the CBF into path planning problems: using gradient-based optimization and using sample-based optimization. A gradient-based optimization programming problem for the nonlinear path planning problems becomes nonlinear optimization, inducing an optimal local solution with excessive computation time. This further results in low performance of the controlled systems and hinders real-time applications.
The sampling-based methods~\cite{lavalle2001randomized, theodorou2010generalized} can address the aforementioned drawbacks. They usually do not require calculating the gradients that might be computationally expensive and impractical for complex systems and finish calculation in a designated time that only depends on sample size.

Model Predictive Path Integral Control (MPPI) algorithm in~\cite{williams2016aggressive} is one of the sample-based algorithms that generates a lot of forward-sampled trajectories to solve the stochastic optimal control problem. By sampling the forward trajectories of the dynamical system, the MPPI algorithm avoids calculating the derivative of the dynamic functions and cost functions~\cite{williams2017information}. Since the forward sampling of random trajectories can be calculated by parallel computing devices, the computation time of the MPPI algorithm is significantly lesser than other traditional methods~\cite{williams2018information}. However, sampling random trajectories in the MPPI algorithm can instigate issues regarding safety guarantees in obstacle-rich environments~\cite{tao2021control}, where most random sample trajectories may violate safety constraints. Also, sample size has a significant influence on the computation time and performance of the sampling algorithm. It remains an open question how the CBF constraints will influence the sample size of the MPPI algorithms. 

In this paper, we formulate a stochastic CBF-MPPI (SCBF-MPPI) algorithm that enhances safety with a probabilistic guarantee for a stochastic system in an obstacle-rich environment. Taking advantage of the MPPI and CBF, the proposed SCBF-MPPI algorithm benefits safety and sampling efficiency compared to the standard MPPI.
In particular, the proposed algorithm improves the sample efficiency by confining sample trajectories in safe regions with changing the variance of the random perturbation. We formally analyze the sampling complexity to show how many sample trajectories required for the given stochastic optimization problem, and to show improved sampling efficiency compared to the standard MPPI. Furthermore, in the simulation with an obstacle-rich environment, we show that the proposed SCBF-MPPI algorithm has better performance in terms of collision avoidance than the MPPI when the same number of samples were used for both algorithms. However, we note that the proposed algorithm induces a sub-optimal solution because the augmentation of the CBF trades off the (infeasible) optimality with enhanced safety.


The rest of the paper is organized as follows. Section \ref{sec:problem} formulates the problem of a stochastic optimization problem with stochastic differential dynamic equations. Section \ref{sec:algorithms} includes the proposed SCBF-MPPI algorithm. Section \ref{sec:sample} uses Chebyshev's Inequality and Hoeffding's Inequality to compare the number of samples required for the SCBF-MPPI algorithm and the MPPI algorithm. Section \ref{sec:simulations} simulates a unicycle robot in a narrow passage environment. Section \ref{sec:conclusion} concludes the algorithm.

\subsection{Related Work}
The augmentation of the CBF in this work is inspired by the previous works on the CBF for fully known and deterministic systems~\cite{ames2019control,ames2016control}. Recent works demonstrate that it is augmentable for uncertain systems~\cite{taylor2020adaptive, zhao2020adaptive, xiao2021adaptive}. In~\cite{taylor2020adaptive}, the authors unify the adaptive control Lyapunov function and adaptive control barrier function to guarantee safety in systems with parametric uncertainties. The paper~\cite{zhao2020adaptive} uses the piece-wise control update law to eliminate the effect of the disturbance. In~\cite{xiao2021adaptive}, the authors introduce adaptive CBF, in the form of a penalty function, to ensure safety for uncertain systems.

Another line of work in safety ensures that control deals with stochastic differential equations instead of parametric uncertainties. In~\cite{koch2019trust}, the authors add chance constraints with the sample-based MPC method for improving safety. In~\cite{pravitra2020}, the authors use $\mathcal{L}_1$ adaptive control augmentation with the MPPI to compensate for the gap between nominal and unknown dynamics. Also, in our previous work~\cite{tao2021control}, we proposed a CBF-based MPPI algorithm that increases the sample efficiency and ensures safety in a nonlinear stochastic path planning problem. However, in the previous work~\cite{tao2021control}, we employed the CBF construction developed for the deterministic systems that are not suitable for stochastic differential equations and did not discuss the influence of the safety guarantee  on the performance or the required sample size.  

In this work, we augment a \emph{stochastic} CBF with the MPPI algorithm for a stochastic guarantee of safety. We also provide a sampling complexity analysis to analyze the influence of the CBF chance constraints. We show that \emph{stochastic} CBF decreases the variance of the random control samples used in the MPPI. In the numerical simulation, the \emph{stochastic} CBF-MPPI has fewer values of the sampling variance, which implies improved sampling complexity by invoking the sampling complexity results in~\cite{yoon2022sample}.    

\section{Problem Statement}\label{sec:problem}
We consider a nonlinear control affine system:
\begin{equation}\label{eq:dynamics}
    dx_t = (f(x_t)+ g(x_t)u_t)dt + \sigma(x_t)dW_t,
\end{equation}
where $x_t \in \mathbb{R}^n$ is the state, $f: \mathbb{R}^n \times \mathbb{R}^n$, $g:\mathbb{R}^n \rightarrow \mathbb{R}^{n\times m}$ and $\sigma: \mathbb{R}^n \rightarrow \mathbb{R}^n$ are locally Lipschitz continuous functions, and $dW_t$ is a Wiener process with $\left < dW_k dW_l \right > = \nu_{kl}(x_t,u_t,t)dt$. We also assume that the stochastic differential equation~\eqref{eq:dynamics} has a strong solution for any control signal $u_t$.

We consider an optimal control problem with a quadratic control cost and a state-dependent cost. The value function $V(x_t,t)$ is then defined as:
\begin{equation}\label{eq:stochastic_prob}
    \min_u \mathbb{E}_{\mathbb{Q}}\left[ \phi(x_T) + \int_t^T(q(x_t,t)+\frac{1}{2}u_t^TR(x_t,t)u_t)dt \right],
\end{equation}
where $\phi(x_T)$ denotes a terminal cost, and $q(x_t,t)$ is a state-dependent cost. $R(x_t,t)$ is a positive definite matrix and needs to satisfy~\cite{theodorou2015nonlinear}
\begin{equation*}
    \nu_{kl}(x_t,u_t,t) = \lambda g(x_t) R^{-1}(x_t,t) g^T(x_t),
\end{equation*}
where $\lambda$ is a constant.

Let $\mathbb{E}_{\mathbb{Q}}$ and $\mathbb{E}_{\mathbb{P}}$ represent the expectations of the trajectories taken with the controlled dynamical system \eqref{eq:dynamics} and the uncontrolled dynamics of the system ($u \equiv 0$ for the dynamical system \eqref{eq:dynamics}).

Let $\mathcal{C}$ represent a specified safe set, which is described by a locally Lipschitz function $h:\mathbb{R}^n\rightarrow \mathbb{R}$ as
\begin{equation*}
    \mathcal{C}=\{x:h(x)\geq 0\}, \qquad \partial \mathcal{C}=\{x:h(x)=0\}.
\end{equation*}

\begin{Pro}\label{pro1}
Given the initial state $x_0 \in \mathcal{C}$, the problem is  to design a control policy that solves the optimization problem defined in \eqref{eq:stochastic_prob} subject to \eqref{eq:dynamics} while guaranteeing $x_t \in \mathcal{C}$ for $\forall t \geq 0$. 
\end{Pro}

\section{SCBF-Constrained MPPI Algorithm }\label{sec:algorithms}
The following section first introduces the SCBF construction and the MPPI sample-based algorithm.  Then we present the SCBF-MPPI algorithm, which has the minimum influence on the sample weights and guarantees safety. Finally, we extend our algorithm to the high degree system.
\subsection{Stochastic Control Barrier Function (SCBF)}\label{sub:cbf}
Guaranteeing the safety of the robots is crucial in the stochastic optimization problem. The CBF algorithms use a Lyapunov-like constraint when the state approaches the boundary of the safe region $\mathcal{C}$ to guarantee safety. However, most CBF algorithms are based on the deterministic system. Yet, solving nonlinear stochastic problems requires considering the uncertainty of the dynamical systems. To resolve the gap between the stochastic and deterministic systems, we use the SCBF, first proposed in~\cite{clark2021control}, using the It\^{o} derivative instead of the Lie derivative to guarantee safety. The use of the It\^{o} derivative adds additional terms in the following definition of SCBF compared to the deterministic one. 

\begin{definition}\cite{clark2021control}\label{def:SCBF}
The function $h:\mathbb{R}^n\rightarrow \mathbb{R}$ is a SCBF for system \eqref{eq:dynamics}, if for all $x$ satisfying $h(x)>0$, there exists $u$ satisfying
\vspace{-0.15cm}
\begin{equation}\label{eq:robust_CBF}
    L_f h(x) + L_g h(x) u + \frac{1}{2} \Tr(\sigma^T \frac{\partial^2 h}{\partial x^2} \sigma) \geq -h(x).
\end{equation}
\end{definition}
\vspace{0.1cm}

Definition \ref{def:SCBF} extends the definition of the CBF to the stochastic system and guarantees safety, as shown in the following theorem.
\begin{The}\cite{clark2021control}
If $u$ satisfies \eqref{eq:robust_CBF} for all time $t$, then $Pr(x_t \in \mathcal{C} \forall t) =1$, provided  $x_0 \in \mathcal{C}$.
\end{The}

The CBF algorithm defines a Quadratic Programming problem that minimizes the difference between the safe control output $u_s$ and the nominal control input $u_n$ and simultaneously satisfies the CBF constraints. We can define the following convex QP problem for the stochastic safe control design:
\begin{equation*}
    \begin{aligned}
&\argmin_{u_s}\quad \frac{1}{2} \left \| u_s - u_n \right \|^2_2 ,\\
&\textrm{s.t.} \quad  L_f h(x) + L_gh(x)u_s + \frac{1}{2}  \Tr(\sigma^T \frac{\partial^2 h}{\partial x^2} \sigma) \geq -\alpha(h(x)).\\
\end{aligned}
\end{equation*}

The SCBF algorithm also averts calculating the derivatives of the dynamic system;  the convex QP optimization problems in the algorithm can be solved in polynomial time~\cite{kozlov1979polynomial}. However, the CBF algorithm can only guarantee the safety of the dynamics. To solve the stochastic path planning problem defined in Problem \ref{pro1}, we will use the MPPI algorithm~\cite{williams2017information}, which also evades the calculation of the derivatives of the dynamic functions.

\subsection{Model Predictive Path Integral Control (MPPI)} \label{sub:MPPI}
To solve the nonlinear stochastic optimization described in problem \ref{pro1}, we apply the MPPI algorithm using Monte Carlo (MC) methods to approximate the optimal control solution. The MPPI algorithm avoids calculating the derivatives of the nonlinear system  or the cost function~\cite{williams2018information}. Since the MPPI algorithm only requires sampling of dynamic trajectories, real-time applications are implemented with the help of parallel computations.

First, the MPPI algorithm samples $K$ trajectories with $T$ being the time horizon. In each trajectory $\tau_i$, let $v_i=[v_{i,0}, \dots, v_{i,T-1}]^T $ be the mean of the control sequence. Let $u_i=[u_{i,0}, \dots, u_{i,T-1}]^T$ be the actual control input sequence. $\epsilon_i=[\epsilon_{i,0}, \dots, \epsilon_{i,T-1}]^T$ represents the disturbance of the control input, and $\epsilon_{i,t} \sim \mathcal{N}(0,\Sigma_{i,t})$.  $[x_{i,0}, \dots, x_{i,T-1}]^T$ denotes the states of the current sample trajectory. 

The iterative update law is:
\begin{equation*}
    u(x_t,t)^*=u(x_t,t)+\frac{\mathbb{E}_{\mathbb{Q}}[\exp (-(1/\lambda) S(\tau)\sigma(x_t) d W(t) ]}{\mathbb{E}_{\mathbb{Q}}[\exp (-(1/\lambda) S(\tau) ]},
\end{equation*}
where $u(x_t,t)$ is the initial control input to be optimized, and $ S(\tau) = \phi(x_T)+ \int_{t_0}^T q(x_t,t)dt$.

The continuous-time trajectories are sampled as a \emph{discretized} system $x_{t+1} =x_t + dx_t$ according to
\begin{equation*}
\begin{aligned}
    dx_t &= \left(f(x_t, t) + g(x_t, t)\vect{u}(x_t, t)\right) \Delta t + \sigma(x_t, t) \delta_t \sqrt{\Delta t}, \\
    &= \tilde{f}(x_t, t) \Delta t + \sigma(x_t, t) \delta_t \sqrt{\Delta t},
\end{aligned}
\end{equation*}
where $\delta_t$ is the Gaussian random vector with independent and identically distributed (i.i.d.) standard normal Gaussian random variables, i.e., $[\delta_t]_i\sim\mathcal{N}(0,1)$, and $\Delta t$ denotes the time step of the time-discretization using Euler–Maruyama method~\cite{platen2010numerical}. Then the discrete-time control update law to approximate the optimal control will be:
\begin{equation}\label{eq:discrete_control}
    u(x_{t_i}, t_i)^*\approx u(x_{t_i}, t_i) + \frac{\sum_{i=0}^{K-1} \exp(-(1/\lambda)\tilde S (\tau_{i,t}))\delta u_{i,t} }{\sum_{i=0}^{K-1} \exp(-(1/\lambda)\tilde S (\tau_{i,t}))},
\end{equation}
where $\delta u_{i,t}=\frac{\delta t}{\sqrt{\Delta t}}$ can be considered as a random control input, and $\tilde{S}(\tau)=\phi(x_{i,t})+\sum_{k=0}^{T-1} \tilde{q}(x_{i,t},v_{i,t}, \epsilon_{i,t})$. The running cost function $\tilde q(x_{i,t}, v_{i,t}, \epsilon_{i,t})$ has the form:

\begin{equation*}
\begin{aligned}
    \tilde{q} (x_{i,t},v_{i,t}, \epsilon_{i,t}) =&q(x_{i,t}) + \frac{1-\nu^{-1}}{2}\epsilon_{i,t}^T R \epsilon_{i,t}, \\
    +&v_{i,t}^T R \epsilon_{i,t} + \frac{1}{2} v_{i,t}^T R v_{i,t} ,
\end{aligned}
\end{equation*}
where $\nu$ is the ratio between the covariance of the injected disturbance $\epsilon_i$ and the covariance of the disturbance of the original dynamics.

The MPPI algorithm solves the nonlinear stochastic problems efficiently based on the dynamic trajectories sampled from normal distributions. However, most sample trajectories may violate the safety constraints and receive a penalty in the reward functions in some extreme environments, such as the obstacle-rich environment. As a result, the sample efficiency of the algorithms becomes poor and eventually influences the performance of the algorithms. Instead of simply using CBF constraints to change the control output, we provide a safe and sampling-efficient algorithm by using SCBF to constrain the sample distributions.

\subsection{SCBF-Based Chance Constrained MPPI Algorithm} \label{sub:MPPI-CBF}
We consider the SCBF-based trust region algorithm, where SCBF constraints are no longer used as a safe filter for the control input. In our previous work~\cite{tao2021control}, we mentioned that if the CBF method compensates for the control output of the MPPI algorithm directly, the exploration of the MPPI algorithm will be hindered. So we design CBF chance constraints to find a trust region for the algorithms to sample. However, the uncertainty in the dynamics is not considered in the CBF chance constraints.  We assume that both the optimal control input and the safe optimal control input can be approximated by Gaussian distribution. We formulate a convex optimization based on the trust region of the SCBF functions as follows:
\begin{equation}\label{eq:chance_cbf}
    \begin{aligned}
    &\argmin_{u} \| u - u_n \| ,\\
    \text{s.t.} &\text{Pr}\left (L_gh(x) u  \geq -h(x) - L_fh(x) -\frac{1}{2} \Tr (\sigma^TH_x\sigma )\right), \\ &\geq 1- \delta ,
    \end{aligned}
\end{equation}
where $u_n$ is a nominal random control input distribution, $H_x =\frac{\partial^2 h}{\partial {x_{i,t}}^2}$ is the hessian matrix of the control barrier function $h$. 

\begin{The}\label{thm:safe}
Let the Gaussian distributions $\mathbb{Q}_0 = \mathcal{N} (\mu_0, \Sigma_0)$ and $\mathbb{Q}_s = \mathcal{N}(\mu_s, \Sigma_s)$ satisfy $\Sigma_0 > \Sigma_s$. For fixed probability $1-\delta$ defined in \eqref{eq:chance_cbf}, if the mean and variance satisfy the constraint:
\begin{equation}\label{eq:chance_constraints}
    A_{i,t} \mu - \alpha A_{i,t} \Sigma A_{i,t}^T \geq b_{i,t},
\end{equation}
where $\alpha$ is the confidence interval  corresponding to the probability $1-\delta$, $A_{i,t} = L_gh(x_{i,t})$, and $b_{i,t} = -h(x_{i,t}) - L_fh(x_{i,t})-\frac{1}{2} \Tr (\sigma^T\frac{\partial^2 h}{\partial {x_t^{i}}^2}\sigma )$, then the sample control input $u_{i,t} \sim \mathbb{Q} (\mu_s, \Sigma_s)$ satisfies the SCBF chance constraints in the optimization problem \eqref{eq:chance_cbf}:
\begin{equation*}
    \text{Pr} \left ( A_{i,t} u \geq b_{i,t} \right ) \geq 1-\delta.
\end{equation*}
\end{The}
\vspace{0.3cm}
\begin{proof}
The left hand side of the inequality can be cast as a Gaussian random variable with mean $A_{i,t}\mu$ and variance to be $A_{i,t} \Sigma A_{i,t}$. Considering the upper bound of the confidence interval for the Gaussian variable, we can simplify the SCBF chance constraint to a linear combination of mean $A_{i,t}\mu$ and variance $A_{i,t} \Sigma A_{i,t}$, which satisfy the inequality $ A_{i,t} \mu \geq  \alpha A_{i,t} \Sigma A_{i,t}^T + b_{i,t}$.
\end{proof}

 Since the variance of the distribution should always be positive semidefinite $\Sigma \succeq 0$, to guarantee the convexity of the previous optimization problem \eqref{eq:chance_cbf}, we reformulate it to:
\begin{equation}\label{eq:SDP_1}
    \begin{aligned}
    \argmin_{\mu, \Sigma}  &\| \mu - \mu_0 \|_1 + \| \Sigma -\Sigma_0 \|_p ,\\
    \text{s.t.}\quad & A_{i,t} \mu - \alpha A_{i,t} \Sigma A_{i,t}^T \geq b_{i,t} ,\\
    &\Sigma \succeq 0.
    \end{aligned}
\end{equation}

\begin{remark}
Suppose that there exists a state $x$ such that the SCBF chance constraint is active. Then  the following inequality holds
\begin{equation*}
    \alpha A_{i,t} \Sigma A_{i,t}^T \leq A_{i,t} \mu -  b_{i,t}.
\end{equation*}
This inequality shows that there is an upper bound on the variance $\Sigma$. Hence, we can assume that $\Sigma_0 \succeq \Sigma$.
\end{remark}

For any positive semidefinite matrices $\Sigma_0, \Sigma$, there exist matrices $P_0, P$ such that $\Sigma_0 = P_0 P_0^T$ and $\Sigma = P P^T$. Using this fact, we can simplify the constraints in \eqref{eq:SDP_1}, and the optimization problem becomes:
\begin{equation}\label{eq:SDP_2}
    \begin{aligned}
    \argmin_{\mu, P} \| \mu - \mu_0 \|_1 + \| P-P_0\|_p, \\
    \text{s.t.} \begin{pmatrix}
      I & \sqrt{\alpha}A_{i,t}P \\
    \sqrt{\alpha} P^T A_{i,t}^T & \mu A_{i,t}-b_{i,t} \\
    \end{pmatrix} \succeq 0.
    \end{aligned}
\end{equation}

The solution to the optimization problem \eqref{eq:SDP_2} provides the safe mean $\mu_s$ and variance $\Sigma_s = PP^T$. In our previous work~\cite{tao2021control}, we proved that the optimization can be simplified to a Semidefinite Programming optimization problem (SDP). In~\cite{helmberg1996interior,warmuth2008randomized,peng2012faster}, it is shown that using parallel computing, the SDP problems can be solved almost as efficiently as linear programming.

Based on the safe mean and variance from SDP optimization \eqref{eq:SDP_2}, we will generate one safe control variation $\delta u_{i,t} \sim \mathcal{N}(\mu_s, \Sigma_s)$ for each sample state $x_{i,t}$ and propagate through discrete dynamics:
\begin{equation}\label{eq:discrete_dynamic}
    x_t^k = x_{t-1}^k + \left( f(x_t^k)+g(x_t^k)(u_{i,t}+\delta u_{i,t} \right).
\end{equation}
We can obtain a sample trajectory $\tau_{i} = \{ x_0, ..., x_T\}$, where $T$ is the time horizon of the MPPI algorithm.

Then we will calculate the cost of the $i^{th}$ sample trajectory by using the cost function $\tilde S(\cdot)$, and using the following equation we will calculate the weight of each trajectory:
\begin{equation}\label{eq:sample_weight}
    \omega_{i} = \exp \left (-\frac{1}{\lambda} \tilde S(\tau_{i}) \right ),
\end{equation}

We use the following control update law:
\begin{equation} \label{eq:control_update}
    u^*(x_{t_i}, t_i) \approx u(x_{t_i}, t_i)+\frac{\sum_{i=1}^K \omega_{i} \delta u_{i,t} }{ \sum_{k=1}^K \omega_{i} }.
\end{equation}

Note that the control update law in \eqref{eq:control_update} cannot guarantee the optimality anymore. The control input $\delta u$ is not a Gaussian variable, and the control update law reaches a sub-optimal solution. So the SCBF-MPPI algorithm's result is more conservative than the original MPPI algorithm.  

We provide the following algorithm:
\begin{algorithm}
\caption{SCBF-MPPI algorithm}
\begin{algorithmic}\label{algo1}
\STATE{\textbf{Given:} $K:$ Number of sample trajectories;}
\STATE{$T:$ Number of timesteps;}
\STATE{$\mu_0,\Sigma_0:$ Initial mean and variance;}
\STATE{$(u_0,u_1, \dots, u_{T-1}):$ Initial control sequence;}
\STATE{$\phi, \tilde{q}, R, \lambda:$ Cost function parameters;}
\STATE{$u_{init}:$ Random initialize control input;}
\WHILE{task is not completed}
    \FOR{$i \leftarrow 0$ to $K-1$}
        \STATE{$x \gets x_0$;}
        \FOR{$t \gets 1$ to $T$}
            \STATE{Solving SDP in \eqref{eq:SDP_2} to get $\mu_s, \Sigma_s$;} 
            \STATE{Generate control variations $\delta u_{i,t}$ $\sim \mathscr{N}(\mu_s,\Sigma_s)$;}
            \STATE{Simulate discrete dynamic \eqref{eq:discrete_dynamic} to obtain $x_{i,t}$;}
            \STATE{Calculate cost function $S(\tau_{i}) \mathrel{+}= \tilde q(x_{i,t},\delta u_{i,t})$;}
        \ENDFOR
        \STATE{Calculate the terminal cost $S(\tau_i) \mathrel{+}= \phi( {x_{i,T}})$}
    \ENDFOR
    \STATE{$\beta \gets \min_i[S(\tau_{i})]$;}
    \STATE{Get sample weights $\omega_{i,t}$  using \eqref{eq:sample_weight};}
    \STATE{Update control input using $\omega_{i,t}$ and $\delta u_{i,t}$ using \eqref{eq:control_update};}
    \STATE{Send $u_{t_0}$ to actuator;} 
    \FOR{$i \gets 0$ to $T-2$}
        \STATE{$u_i = u_{i+1}$;}
    \ENDFOR
    \STATE{$u_{N-1}=u_{init}$;}
\ENDWHILE
\end{algorithmic}
\end{algorithm}

\subsection{SCBF-Based Chance Constrained MPPI Algorithm for High Relative Degree System} \label{sub:high-order}
The CBF chance constraints may fail with  certain states $x$ when $\frac{\partial h}{\partial x}g(x)=0$. The safe constraints will become infeasible or allow dangerous control input. To satisfy the safety constraints of high-order system, we have the following definition of $h^r(x)$ from previous work~\cite{clark2021control} for $r=0, 1, \dots$, as $h^0(x) = h(x)$ and
\begin{equation*}
    h^{r+1}(x) = \frac{\partial h^r}{\partial x}f(x) + \frac{1}{2}\Tr\left ( \sigma^T \left( \frac{\partial^2 h^r}{\partial x^2} \right )  \sigma\right ) + h^r(x).
\end{equation*}

Define $\mathcal{C}^r=\{ x:h^r(x) \geq 0 \}$, and the intersection of these sets is defined as $\bar {\mathcal{C}}^n = \bigcap^n_{r=0} \mathcal{C}^r$. We provide the high-order chance SCBF optimization problem:
\begin{equation}\label{eq:chance_high_cbf}
    \begin{aligned}
    &\argmin_{u} \| u - u_n \| \\
    \text{s.t.} &\text{Pr}\left (\frac{\partial h^r}{\partial x}g(x) u  \geq -h^r(x) - \frac{\partial h^r}{\partial x} f(x) -\frac{1}{2} \Tr (\sigma^TH^r_x\sigma )\right) \\ &\geq 1- \delta ,
    \end{aligned}
\end{equation}
where $H^r_x = \frac{\partial^2 h^r}{\partial x^2}$. With the previous definition, we have the following theorem.

\begin{The}\label{thm:high_order}
Let the Gaussian distributions $\mathbb{Q}_0$ $ = \mathcal{N} (\mu_0, \Sigma_0)$ and $\mathbb{Q}_s = \mathcal{N}(\mu_s, \Sigma_s)$ satisfy $\Sigma_0 > \Sigma_s$. For a $n$-degree high order stochastic system, we assume that the current state $x\in \bar{\mathcal{C}}^r$ satisfies the condition  $\frac{\partial h^r}{\partial x} g(x) \geq 0$ for $r <n$. Then for fixed probability $1-\delta$, if the mean and variance satisfy the constraint:
\begin{equation}\label{eq:high_cbf}
    A^r \mu_s - \alpha A^r \Sigma {A^r}^T\geq b^r,
\end{equation}
where $A^r = \frac{\partial h^r}{\partial x} g(x)$, $b^r = -\frac{\partial h^r}{\partial x} f(x) -\frac{1}{2} \left ( \sigma^T \frac{\partial^2 h^r}{\partial x^2} \sigma \right ) -h_r(x) $,  the sample control $u \sim \mathbb{Q}(\mu_s, \Sigma_s)$ will guarantee the safety defined by the chance constraints in \eqref{eq:chance_high_cbf}.
\end{The}
\vspace{0.2cm}
\begin{proof}
Suppose that $\mu_s, \Sigma_s$ satisfying the conditions of the theorem are chosen at each time step t. By theorem \ref{thm:safe}, the ineqaulity in \eqref{eq:high_cbf} implies that $\text{Pr} (h^n(x_t) \geq 0) < 1 - \delta$. By definition of function $h^r(x)$ and the assumption that $\frac{\partial h^{n-1}}{\partial x} g(x) u \geq 0$, we also have $\text{Pr} (h^{n-1}(x_t) \geq 0) < 1 - \delta$ for all time $t$. Proceeding inductively, we then have $\text{Pr} (h^r(x_t) \geq 0) < 1 - \delta$ for all $r=0, \dots, n$, and hence in particular $\text{Pr}( h(x_t) = h_0(x_t) \geq 0) \geq 1 -\delta$ for all t.
\end{proof}
\vspace{0.1cm}

We propose the following high-order SDP optimization problem based on Theorem \ref{thm:high_order}:
\begin{equation*}
    \begin{aligned}
    \argmin_{\mu, \Sigma}  &\| \mu - \mu_0 \|_1 + \| \Sigma -\Sigma_0 \|_p \\
    \text{s.t.}\quad & A^r_{i,t} \mu - \alpha A^r_{i,t} \Sigma {A^r_{i,t}}^T \geq b^r_{i,t} \\
    &\Sigma \succeq 0.
    \end{aligned}
\end{equation*}

\section{Sample Size Analysis}\label{sec:sample}

The number of sample trajectories has a significant effect on the performance and the computation time of the sampling-based algorithms. With a large sample size, the performance of the algorithm will be better, but the computation time of the algorithm will increase significantly as well. Intuitively, the SCBF-MPPI algorithm provides a restricted but safe region for sample trajectories which should help increase the sample efficiency. Our previous work on the sampling complexity of the PI method~\cite{yoon2022sample} uses Hoeffding's inequality and Chebyshev's inequality. We will provide a sampling complexity analysis of this intuitive idea. We discuss the case of one dimensional control input $[\delta u_{i,t}]_i \sim \mathcal{N}([\mu_t]_i, [\Sigma_t]_i)$ in this paper, and a similar result can be extended to high dimensional control input straightforwardly.

\begin{ass}\label{ass:1}
Assume that the error bound $\epsilon_1$ of the Chebyshev's inequality is smaller than the expectation of $\omega$
\vspace{-0.1cm}
\begin{equation*}
    \epsilon_1 < \E [\exp (-\frac{1}{\lambda} S(\tau_i)].
\end{equation*}
\end{ass}
\vspace{0.1cm}

\begin{ass}\label{ass:cost}
We suppose that the running cost function $\tilde q(x_{i,t})$ and terminal cost function $\phi(x_{i,T})$ are quadratic functions. 
\end{ass}
\vspace{0.1cm}
\begin{The}
Under Assumptions \ref{ass:1} and \ref{ass:cost}, the number of sample $N$ of the original MPPI control update law defined in \eqref{eq:discrete_control} is larger than the number of sample $N^s$ of the control update law of the SCBF-MPPI algorithm \eqref{eq:control_update} given the same sampling complexity error bound $\epsilon$ and risk probability $\rho$. 
\end{The}
\vspace{0.1cm}
To prove the above theorem we have the following lemmas and propositions.
\vspace{0.1cm}
\begin{lemma}\label{lemma:1}
For any random variables $X, Y$, we have:
\vspace{-0.1cm}
\begin{equation*}
    \Var \left [ XY \right]   \leq 2\Var[X] \Var[Y] + 2 \Var [Y] \E[X]^2 .
\end{equation*}
\end{lemma}
\vspace{0.1cm}
\begin{proof}
For any random variables we have that $\Var[A+B] \leq \Var [A] + \Var[B]$. Then consider $A = (X-\E[X])Y$ and $B = \E [X] Y$; so we have:
\begin{align*}
    \Var [XY] &= \Var \left [ (X-\E[X])Y +\E [X] Y \right ] \\
    & \leq 2 \Var \left [ (X-\E [X]) Y \right] + 2 \Var \left [ \E [X] Y \right ] \\
    & = 2\E \left [ (X - \E[X] )^2 Y ^2 \right ] - \E \left [(X-\E[X])Y \right ]^2 \\
    & + 2 \Var \left[\E [X] Y \right ] \\
    & \leq 2\E \left [ (X - \E[X] )^2 Y ^2 \right ] + 2 \Var \left[\E [X] Y \right ] \\
    & \leq 2\E \left [ (X - \E[X] )^2  \right ] \E [Y^2] + 2 \Var \left[\E [X] Y \right ] \\
    &= 2 \Var [X] \E[Y^2] + 2\Var[Y] \E[X]^2 \\
    &= 2\Var[X] \Var[Y] -2\Var[X] \E[Y]^2 \\
    &+ 2 \Var [Y] \E[X]^2 \\
    & \leq 2\Var[X] \Var[Y] + 2 \Var [Y] \E[X]^2.
\end{align*}
\end{proof}

\begin{lemma}
It holds that
$\Var [\omega] \leq (1-\E [\omega]) \E [\omega] \leq \E [\omega] \leq 1$.
\end{lemma}
\vspace{0.2cm}
\begin{proof}
Since $\omega = \exp(-\frac{S(\tau)}{\lambda})$ and since the cost-to-go function $S(\tau) \geq 0$ by assumption \ref{ass:cost}, then $\omega \in [0,1]$ is a bounded random variable and its variance is also bounded.
\end{proof}

\begin{corollary}\label{cor:N_1}
It holds that
\begin{equation}\label{eq:N1}
    \mathbb{P} \{ | \hat E_1 - \E [\omega] | \geq \epsilon_1 \} \leq \rho_1 \coloneqq 2\exp(-N_1\epsilon_1^2),
\end{equation}
where $\hat E_1 = \sum_{i=1}^T \exp(-\frac{1}{\lambda} S(\tau_i)) $, $\epsilon_1$ is the error bound of the discrete sample estimate, and $\rho_1$ is referred to as risk probability of not satisfying the error bound. The number of sample $N_1$ can be calculated as:
\begin{equation}\label{eq:N_1}
    N_1 = -\frac{1}{\epsilon_1^2}  \log{\frac{\rho_1}{2}}.
\end{equation}
\end{corollary}
\vspace{0.2cm}
\begin{proof}
Since the $\omega \in [0,1]$, then by Hoeffding's inequality:
\begin{align*}
   \mathbb{P} \{ | \hat E_1 - \E [\omega] | \geq \epsilon_1 \} &\leq 2 \exp \left ( -\frac{N_1\epsilon_1^2}{(\omega_{\max}-\omega_{\min})^2} \right )
   \\&\leq 2\exp(-N_1\epsilon_1^2) .
\end{align*}
\end{proof}

\begin{lemma}\label{lemma:sample_number}
Let $\hat E_2$ denote the resample control update $\hat E_2 \coloneqq \frac{1}{N} \sum^N_i (\frac{\omega_i \delta u_i}{\E [\omega]})$ and $E_2 = \mathbb{E} \left [ \frac{\omega \delta u}{\mathbb{E}[\omega]} \right ] $. We have the following error bound $\epsilon_2$ and the number of sample  $N_2$
\begin{equation} \label{eq:N2}
    \mathbb{P}\left \{ \left | \left [ \hat E_2-E_2\right ]_i \right |\geq \epsilon_2 \right \}  \leq \rho_2 \coloneqq \frac{\Gamma}{N_2\epsilon_2^2}\left (\exp\left(\frac{2\mathbb{E}[S(\tau)]}{\lambda}\right)\right ),
\end{equation}
where $\Gamma = 4 \Var [\delta u]$. 
\end{lemma}
\vspace{0.2cm}
\begin{proof}
First we give the bound of the variance of the control input $E_2$
\begin{align*}
    \Var \left [ \frac{\omega  \delta u_t}{\E [\omega]} \right ] &= \frac{1}{\E [\omega]^2} \Var \left [ \omega[\delta u_t] \right ] \\
    &\leq \frac{2 \Var[\omega] \Var[\delta u] + 2 \Var [\delta u]}{\E [\omega]^2} \\
    &\leq \frac{2 \Var[\omega] \Var[\delta u] + 2 \Var [\delta u]}{ \left (\exp \left(-\frac{\E [S(\tau)]}{\lambda}\right)\right)^2} \\
    &\leq \frac{4 \Var [\delta u]}{\left (\exp \left(-\frac{\E [S(\tau)]}{\lambda}\right)\right)^2} ,
\end{align*}
where the first inequality uses the result of Lemma \ref{lemma:1},  and since $\omega = \exp (-\frac{1}{\lambda} S(\tau) )$, then we can obtain that $\omega \in (0,1)$. The second inequality follows from Jensen's inequality with convex exponential equation:
\begin{equation*}
    \exp \left (-\frac{\E [S(\tau)]}{\lambda} \right) \leq \E \left [ \exp \left ( \frac{S(\tau)}{\lambda} \right )\right ] = \E[\omega].
\end{equation*}
By using  Chebyshev's inequality we have:
\begin{align*}
        &\mathbb{P} \left \{ \left | E_2 - \hat E_2 \right | \geq \epsilon \right \} \coloneqq \frac{\text{Var}[E_2]}{N_2 \epsilon_2^2} .
\end{align*}
\end{proof}

\begin{corollary}\label{cor:N_2}
Under Assumptions \ref{ass:1} and \ref{ass:cost}, the MC error bound in \eqref{eq:N2} becomes
\begin{equation*}
    \mathbb{P}\left \{ \left | \left [ \hat E_2-E_2\right ]_i \right |\geq \epsilon_2 \right \}  \leq  \frac{\Gamma}{N_2\epsilon_2^2}\left (\frac{1}{\hat E_1 - \epsilon_1}\right )^2,
\end{equation*}
where $\epsilon_1$ is the first MC error bound from \eqref{eq:N1}. Then we conclude that for the error bound $\epsilon_2$ and the risk probability $\rho_2$, the number of sample $N_2$ can be calculated:
\begin{equation}\label{eq:N_2}
    N_2 = \frac{4 \Var[{\delta u}]}{\rho_2\epsilon_2^2}\left (\frac{1}{\hat E_1 - \epsilon_1}\right )^2.
\end{equation}
\end{corollary}
\vspace{0.3cm}
\begin{proof}
From the proof in  Lemma \ref{lemma:sample_number}, we have
\begin{equation*}
    \Var \left [ \frac{\omega  \delta u_t}{\E [\omega]} \right ] \leq \frac{4 \Var[\delta u]}{\E [\omega]^2}.
\end{equation*}
Using the inequality $| \hat E_1 - \E [\omega] | \geq \epsilon_1$, we have the following relation for $(\E[\omega])^2$:
\begin{equation*}
    \frac{1}{(\hat E_1+\epsilon_1)^2} \leq \frac{1}{(\E[\omega])^2} \leq \frac{1}{(\hat E_1-\epsilon_1)^2}.
\end{equation*}
Then the above follows using the Chebyshev's inequality.
\end{proof}

Since the control output distribution of the MPPI algorithm $\delta u$ has greater variance than the safe control output distribution of the SCBF-MPPI algorithm $\delta u_s$, by using the conclusion of Corollary \ref{cor:N_2}, we can conclude that to reach the same error bound $\epsilon_2$ and risk probability $\rho_2$, the number of sample of the MPPI algorithm $N_2$ is greater than the number of SCBF-MPPI algorithm $N_2^s$. For the same error bound $\epsilon_1$ and risk probability $\rho_1$, the number of sample of both algorithms are the same based on the Corollary \ref{cor:N_1}. So the required sample  number for the MPPI algorithm $N=\max(N_1,N_2)$ is also greater or equal to the number of sample for the SCBF-MPPI algorithm $N^s = \max(N_1, N_2^s)$.

\section{Simulations}\label{sec:simulations}
This section implements the SCBF-MPPI algorithm and MPPI algorithm on a stochastic nonlinear control affine system. We compare the performance of both algorithms with different sample sizes and plot  the sample trajectories. We also calculate the sample size for the fixed error bound.
\subsection{Unicycle Dynamics}
We implement our algorithms on a two-dimensional unicycle dynamical system with:
\begin{equation*}
    \begin{bmatrix}
\dot {x}\\ 
\dot {y}\\ 
\dot {\theta}
\end{bmatrix} = \begin{bmatrix}
\cos \theta & 0\\ 
\sin \theta & 0\\ 
0 & 1
\end{bmatrix}\begin{bmatrix}
v \\ 
\omega 
\end{bmatrix} + \sigma dW(t),
\end{equation*}
where $x, y$ are the coordinates, $\theta$ is the angle, $v$ is the linear velocity,  $\omega$ is the angular velocity of the unicycle model, $\sigma$ is the identity matrix, and $W(t)$ is a Brownian motion. The time step for the discrete-time simulation is $\Delta t = 0.05 s$.

\subsection{Simulation Setup}
We demand a probability of $1-\gamma = 0.997$ of avoiding all obstacles. The parameters for the sampling algorithm are set: the time horizon $T =20$, and $\lambda =1$. 

We consider a narrow passage environment where the safe set $\mathcal{C}$ is defined as $\mathcal{C}:\left \{(x,y) | \sin \frac{\pi}{2} x < y < \sin \frac{\pi}{2} x + \alpha \right \}$, where $\alpha =1$ is the width of the narrow passage. The control barrier functions are:
\begin{equation*}
    \begin{aligned}
    h_1 &= y - \sin x > 0, \\
    h_2 &= \sin x +\alpha - y >0.
    \end{aligned}
\end{equation*}

We consider a stochastic path planning problem where the start position of the unicycle is $X_0 = [x_0, y_0]^T = [0, 0.5]^T$. The goal is set as $X_g=[x_g, y_g]^T = [4, 0.5]^T$. The initial angle state $\theta_0=0$, and the running cost function is
\begin{equation*}
    q(x,y,\theta) = \| X- X_g\|_2^2 + 1000 * \mathds{1}_{X\in \bar {\mathcal{C}}},
\end{equation*}
where $\bar {\mathcal{C}}$ is the complementary set to the safe set $\mathcal{C}$ over $\mathbb{R}^2$, and $\mathds{1}$ is the indicator function. 
\begin{figure}[h]
    \centering
    \includegraphics[width=\linewidth]{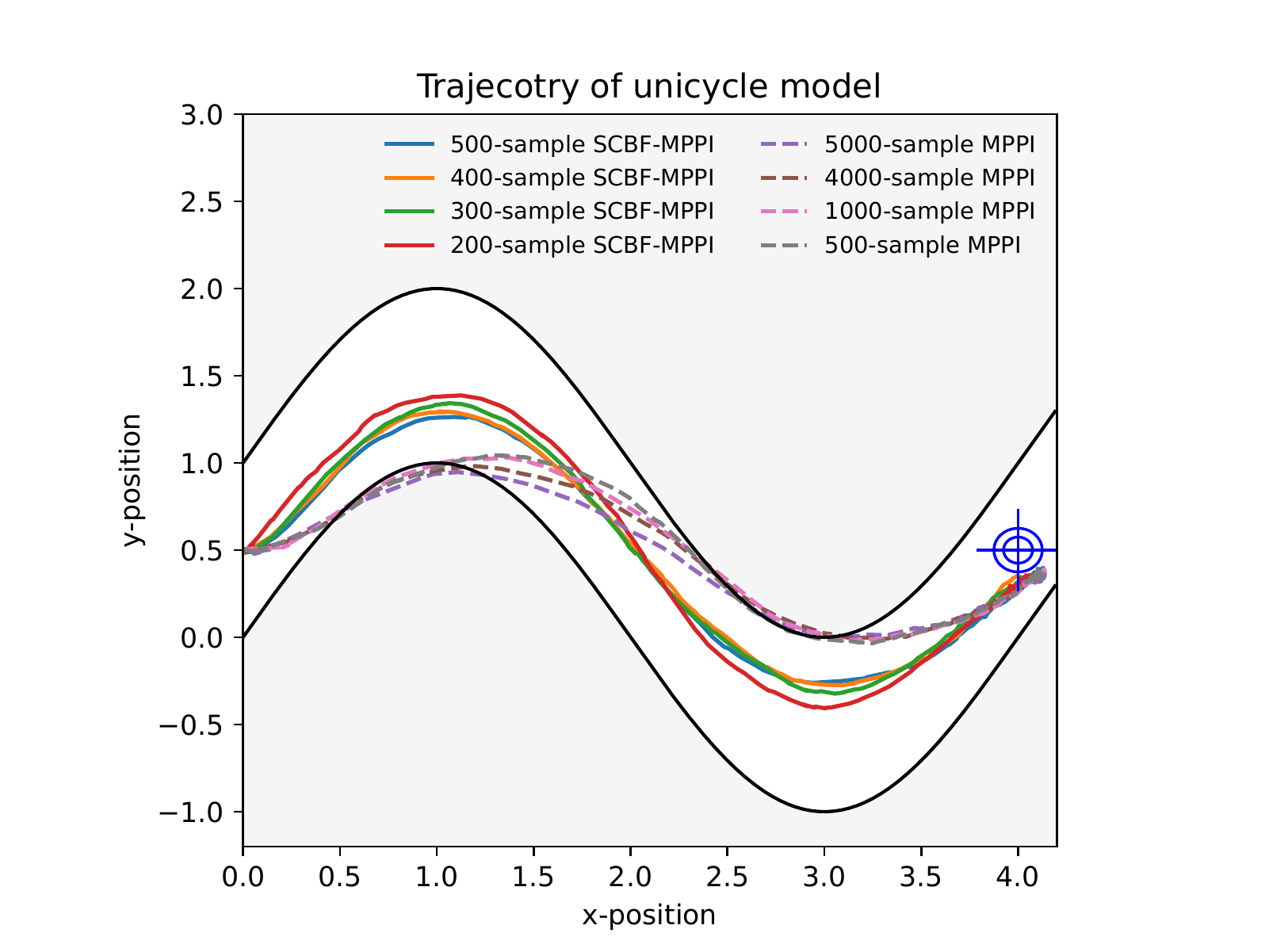}
    \caption{Path planning result with different sample sizes.}
    \label{fig:sin_plots}\vspace{-0.2cm}
\end{figure}

\subsection{Results}
\begin{figure}[h]
    \centering
    \includegraphics[width=0.45\textwidth]{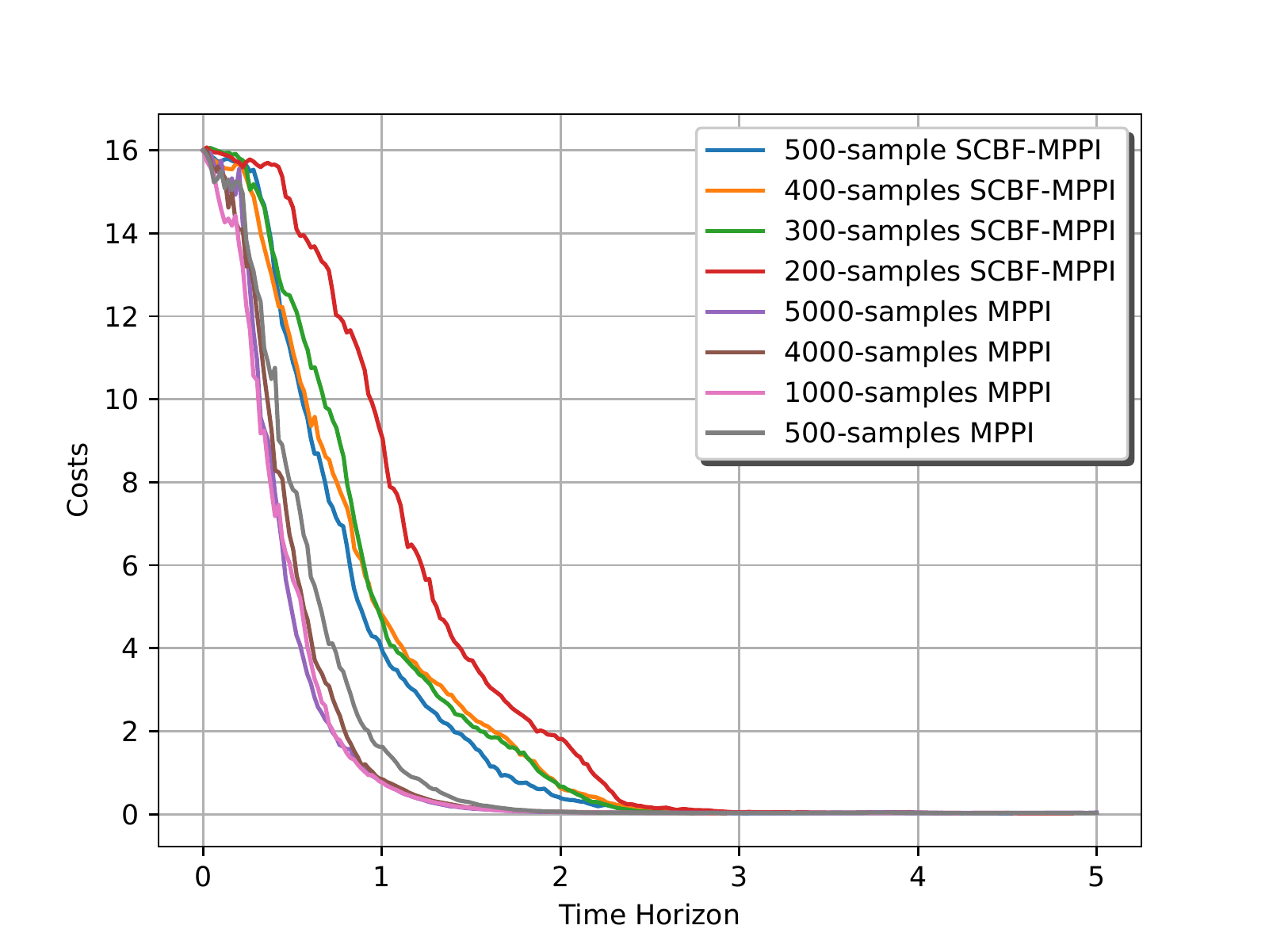}
    \caption{Cost with different sample sizes.}
    \label{fig:cost_plots}
\end{figure}

Figure \ref{fig:sin_plots} illustrates a result for the unicycle robots navigating through a narrow passage for at most 12.5 seconds. The black line represents the boundary of the safe set. The blue cross represents the target position. We first implemented the MPPI algorithm with 500, 1000, 4000 and 5000 samples. The robotic systems can reach the target position but may violate safety in some states. Then, we implement the SCBF-MPPI algorithm with 200, 300, 400, and 500 samples where the safety is guaranteed and also reaches the target successfully. However, the SCBF-MPPI algorithm behaves more conservatively and spends more time steps for reaching the goal. In Figure \ref{fig:cost_plots}, we plot different algorithms' running cost function values with varying sample sizes. The SCBF-MPPI algorithm converges slower than the MPPI algorithm, which illustrates the same result in Figure \ref{fig:sin_plots}: the SCBF-MPPI algorithm is more conservative. The CBF constrained sample distribution causes the conservativeness of the control output.

We define the collision rate as how many states in the trajectory violate the safety constraint. We further define the average time to finish (TTF) to be the number of time steps to reach a vicinity area of the target. In this experiment, we define the vicinity area is a circle with a radius of $0.15$. We repeat both algorithms ten times to calculate the average collision rate and average time steps. Table \ref{tab:table1} shows that in both the MPPI and the SCBF-MPPI algorithms, the average TTF will decrease when the sample size is larger, which means the performance will be better. The MPPI algorithm has a lower average TTF than the SCBF-MPPI, which indicates that SCBF-MPPI is more conservative. However, the collision rate of SCBF-MPPI remains 0, which implies the safety constraints are always satisfied during the experiments.

\begin{figure}[t]
  \centering
  \captionsetup{width=0.9\linewidth}
    \includegraphics[width=0.45\textwidth]{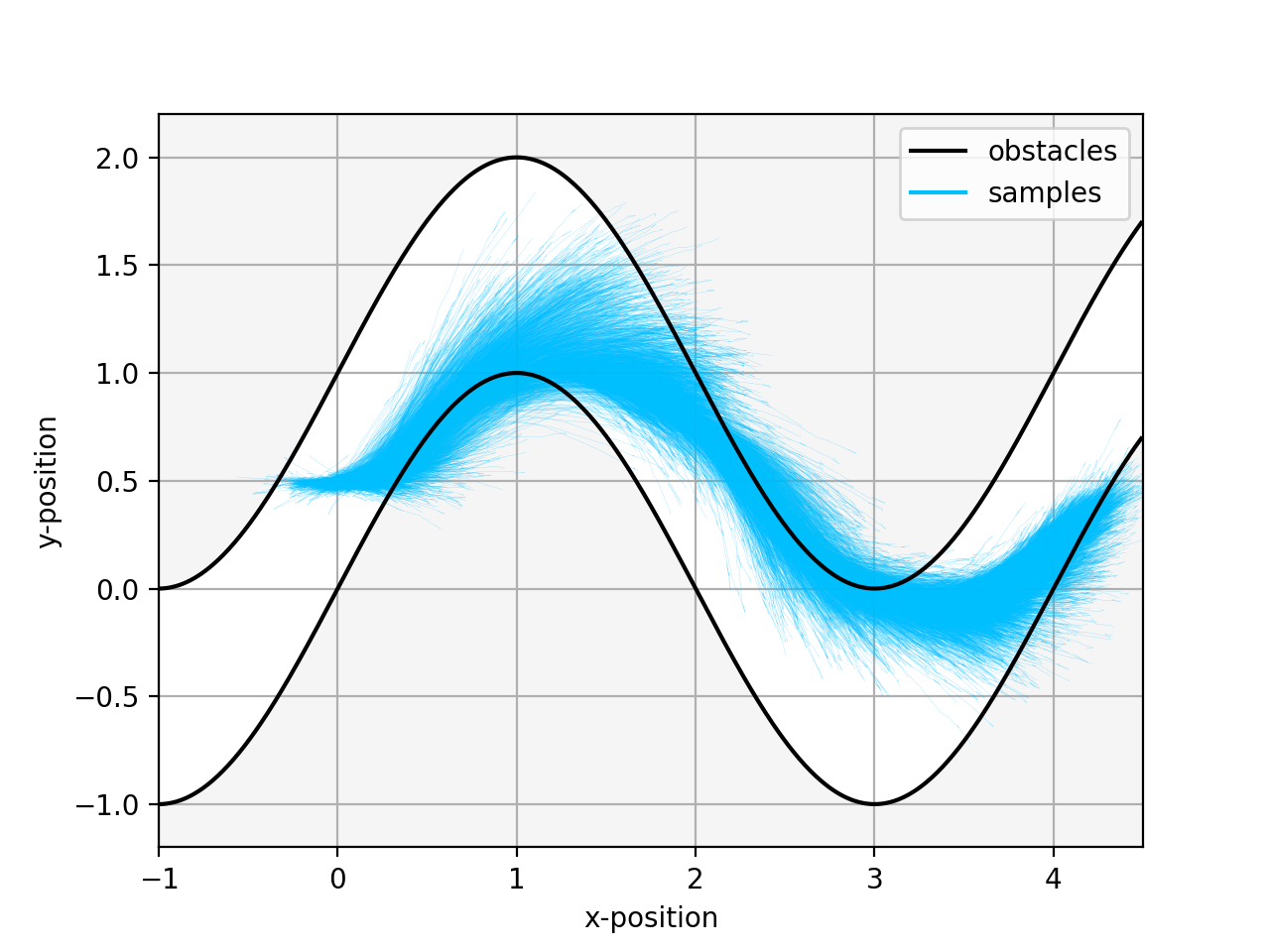}
    \caption{The sample trajectories of the MPPI algorithm.}
    \label{fig:MPPI_sample}
\end{figure}

\begin{figure}
  \centering
  \captionsetup{width=0.9\linewidth}
    \includegraphics[width=0.45\textwidth]{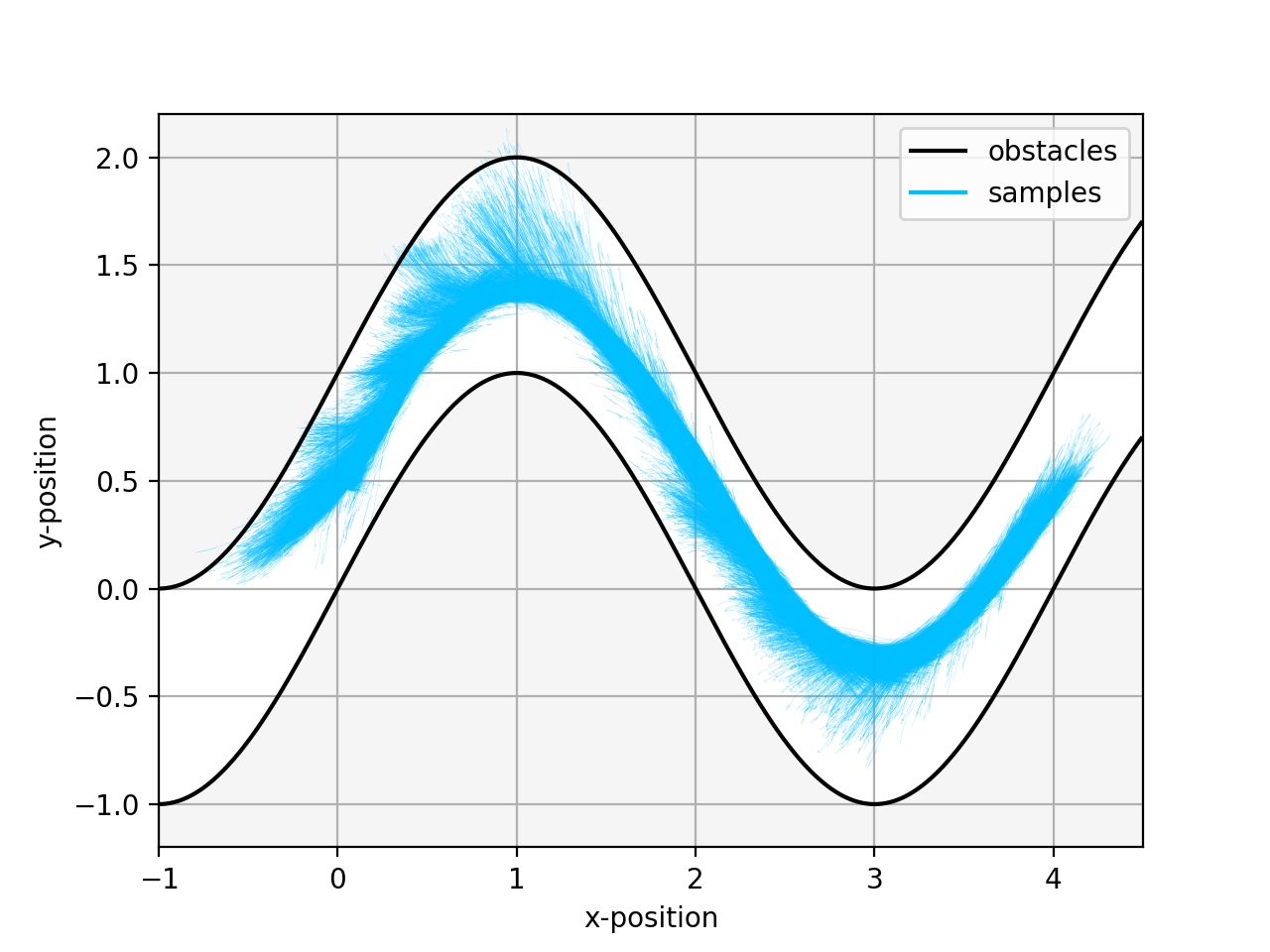}
    \caption{The sample trajectories of the SCBF-MPPI algorithm.}
    \label{fig:MPPI_CBF_sample}
\end{figure}

In Figure \ref{fig:MPPI_sample} and Figure \ref{fig:MPPI_CBF_sample}, we plot all the sample trajectories of the MPPI algorithm and the SCBF-MPPI algorithm when the sample size is 200. The sample trajectories are in blue color, and the boundary of the safe set is in black color. The sample trajectories in the MPPI algorithm follow a Gaussian random variable with fixed mean and variance. For the SCBF-MPPI algorithm, the mean and variance of the sample trajectory are obtained based on the SDP optimization. As a result, all sample trajectories are in the safe set. Note that some aggressive and dangerous sample trajectories in the SCBF-MPPI algorithm will be turned opposite because of the confidence interval we set. However, these samples will have a relatively low cost and have minor effect on the final control output.

\begin{table}[t]
    \centering
    \caption{
    MPPI and CBF-MPPI of different sample}
    \begin{tabular}{c c c}
    \toprule
         Case  & Collision Rate & Avg. TTF  \\ \midrule \midrule
         200 samples MPPI  &0.0454 & 140.4\\ 
         500 samples MPPI &0.0294 & 129.6 \\   
         200 samples SCBF-MPPI & 0.0& 163.6 \\
         500 samples SCBF-MPPI & 0.0 & 156.1 \\ 
         \midrule
    \end{tabular}
    \label{tab:table1}\vspace{-0.2cm}
\end{table}

We sample 500 trajectories in our experiments, and set the desired bound $\epsilon_1 = 0.05, \epsilon_2 =0.1,$, the allowable risk of failure $\rho_1 = 0.05,  \rho_2 = 0.1$. We calculate the numbers of sample $N_1$ and $N_2$ based on the equations \eqref{eq:N_1} and \eqref{eq:N_2} at time step $T=50$. Table \ref{tab:table2}  In conclusion, it shows that the CBF-MPPI algorithm needs smaller sample size than the MPPI algorithm. 

\begin{table}[ht]
    \centering
    \caption{
    Sampling number for $N_1$ and $N_2$}
    \begin{tabular}{c c c c}
    \toprule
         Algorithm  & T& $N_1$ &$N_2$ \\ \midrule \midrule
          MPPI  & 50& 1476 & 2973\\ \midrule
        SCBF-MPPI & 50 & 1476 & 584 \\
         \midrule
    \end{tabular}
    \label{tab:table2}
\end{table}
\vspace{-0.2cm}

\section{Conclusion and Discussion}\label{sec:conclusion}
We propose a SCBF-MPPI algorithm that utilizes the safe SCBF constraints to determine the mean and variance of the random control trajectories for calculating the path-integral control. The augmentation of the SCBF improves safety as it shows zero collision rate in the simulations of the unicycle model. The augmentation will defer the control from the (potentially infeasible) optimal solution provided by the MPPI. The improved safety can also indirectly be observed through the fewer number of sample required by the SCBF-MPPI than the MPPI for the same level of assurance. Using sampling complexity analysis and simulations, we show that the CBF-MPPI algorithm needs fewer samples due to the smaller variance of the control distribution. In the simulations, we formulate a narrow passage problem with the unicycle model to show the safety enhancement and sample efficiency of the CBF-MPPI algorithm.

\bibliographystyle{ieeetr}
\bibliography{citation}

\end{document}